\documentclass[11pt]{article}
\usepackage{graphicx} % Required for inserting images
\usepackage{xcolor}
\usepackage{natbib}
\usepackage{amsmath}
\usepackage{amssymb}
\usepackage{bbm}
\usepackage[figuresright]{rotating}
\usepackage{multicol,ragged2e}
\usepackage{setspace}
\usepackage{algorithm}
\usepackage{algpseudocode}
\usepackage{authblk} % author affiliations

 \newcounter{acounter}
\newcommand\startsupplement{%
    \makeatletter 
       \setcounter{table}{0}
       \renewcommand{\thetable}{S\arabic{table}}   
       \setcounter{figure}{0}
       \renewcommand{\thefigure}{S\arabic{figure}}   
    \makeatother}

\title{Methods to address measurement error in both Outcome and Covariates}
\author[1,2,*]{Pamela~A. Shaw}
\author[3]{Bryan~E. Shepherd}
\affil[1]{Biostatistics Division, Kaiser Permanente Washington Health Research Institute, Seattle, WA, USA}
\affil[2]{Department of Biostatistics, University of Washington, Seattle, WA, USA}
\affil[3]{Department of Biostatistics, Vanderbilt University, Nashville, Tennessee, USA}
\affil[*]{Corresponding author: Pamela Shaw, pamela.a.shaw@kp.org}
\date{\today}

\begin{document}

\maketitle

\section{Introduction}

Measurement error is a problem that arises in many settings in biostatistical applications, where exposures and outcomes can be complex and difficult to measure accurately. Data from large observational cohort studies can be particularly error-prone due to the impracticality of precise measurement on a large number of individuals.  Increasingly, many clinical studies are leveraging data from large administrative databases, such as Medicare, or electronic health records (EHR) data as a cost-effective way of collecting a rich set of health-related outcomes and exposures.   In study variables derived from administrative or EHR data, errors arise from two fundamental limitations: 1) the data were not collected to support research and 2) study variables are typically derived from automated computer algorithms that may miss or falsely interpret the EHR data. For example, audits of routinely collected data from cohorts of individuals living with HIV have discovered error rates as high as 56\% for key variables \citep{duda12, lotspeich2022}. In this and other settings reliant on EHR data, errors can arise in both outcomes and exposures. In particular, error in one variable can set off a cascade of correlated errors. For example, if there was an error in the date of diagnosis, baseline covariates meant to represent the patient’s health at time of diagnoses would then be from the wrong time and time from diagnosis to event outcomes would also be incorrect. 

It has been well established that errors in exposures can bias regression coefficients in association analyses \citep{carroll06,yi17,keogh20,shaw20}. Research has also shown errors in outcomes, in both linear and non-linear models, can bias analyses -- particularly when errors are known to be covariate-dependent \citep{keogh20,shaw20}. Most of the measurement error literature to date, however, has focused on errors in either the exposure or the outcome, but not both. Even in cases where analytical approaches have been well-established \citep{carroll06}, such as in the simpler case of errors only in covariates, researchers often ignore the errors in their analyses or make incorrect claims about the impact of the measurement error on their study results \citep{brakenhoff2018measurement,shaw2018}. 

In this tutorial, we will discuss methods of analysis for settings where there are errors potentially in both outcomes and exposures and illustrate their use in practical examples. In these examples, we will compare results across the methods and discuss the relative advantages of the different approaches for various settings. Importantly methods to address measurement error generally need validation or replication data that provide information on the structure of the measurement error. We will consider methods of analysis that can be applied to understand the potential impact of measurement error in this setting.    

In section 2, we describe the data structure for the setting of interest and introduce notation. In section 3, we present a motivating example, and we describe a synthetic dataset we generate from this example which we will use to illustrate the methods; the synthetic dataset is available to readers so that they can reproduce the results in this paper. We then describe several methods available to address the measurement error in both outcomes and covariates and illustrate their application in Section 4. In section 5, we discuss methods that can be applied when there are no available data to inform the measurement error structure. Finally, we conclude with a discussion of some general observations and guidance in Section 6.

\section{Data structure and notation}

Generally to assess, and therefore account for,  the measurement error, additional data that are informative of the measurement error structure are needed. In a two-phase design, data available for the entire cohort are collected at phase 1, and additional data are obtained on a subset at phase 2 . For example, error-prone data can often be extracted relatively inexpensively from an EHR database using computer algorithms in phase 1, and in phase 2, higher quality data are obtained by a labor intensive chart review on a subset. When the data collected in phase 2 are considered error-free, or the gold standard, this subset is often referred to as the \textit{validation subset}. In many settings, the data collected at phase 2 may still be error-prone, but of higher quality and potentially useful for learning about the measurement error. For example, in nutritional epidemiology setting, self-report is an inexpensive way to collect dietary exposures that can be subject to both systematic and random (unbiased) error; whereas, for some nutrients there are biomarkers that  measure intake with only random error. Settings where the data at phase 2 have unbiased (classical) measurement error are typically referred to as a calibration subset, a name that refers to a common method to address measurement error, namely regression calibration. 

We denote error-prone versions of variables with a star. Thus, $X^*$ and $Y^*$ denote the error prone covariate vector and outcome of interest, respectively, and $X$ and $Y$ denote the error-free versions of these variables. %In settings where both the phase 1 and phase 2 variables are prone to error, we use a double star to denote the more precise, but still error-prone, phase 2 data, e.g. $(X^{**}, Y^{**})$ are available on the calibration subset. 
In time-to-event settings, $Y$ may be the pair of variables $(T, \Delta)$, where $T$ is the censored event time and $\Delta$ is the observed event indicator. These variables may also be in error, denoted as $(T^*, \Delta^*)$. In some settings there may also be a vector $Z$ of precisely obtained variables at phase 1. 

In the examples that follow, it will also be useful to define the phase 2 indicator $R$ and distinguish between exposures in the outcome model $X$ and auxiliary variables $A$, which may be useful for estimation, but do not appear in the outcome model.  Let $N$ and $n$ be the size of the phase 1 and phase 2 samples, respectively. While we do not formally address regularity conditions necessary for the discussed methods to have desirable properties, an assumption that is generally sufficient regarding the size of the validation subset is that $n/N$ converges to a constant $\rho > 0$.

\section{Motivating data example}

\subsection{Maternal Weight Gain (MWG) Study} 

As a running example, we consider a study looking at maternal weight gain, childhood obesity, and childhood asthma. \citet{shepherd23} obtained EHR data for 10,335 mother-child dyads receiving care at Vanderbilt University Medical Center, where all mothers were distinct. Data for mothers were collected up to one year prior to conception to the date of delivery, and data for newborn children were collected from birth until age six years. Maternal weight gain during pregnancy and maternal body mass index (BMI) at conception were estimated using functional principal components analyses. Other variables collected for mothers included age at delivery, race, ethnicity, diabetes status, depression, insurance status, smoking history, and asthma diagnosis. Variables collected for children included weight, sex, asthma status, and an indicator that they were delivered using a Cesarean section.  Because of concerns regarding the quality of EHR data, and a few key variables only being available from chart review (e.g., estimated conception date), a research nurse performed extensive chart reviews of the data for 996 mother-child dyads. These dyads were chosen using stratified random sampling that occurred over multiple sampling waves where information used from the prior waves of chart reviews were used to target the most informative records for chart review in subsequent waves. The chart reviewed data are assumed to be correct throughout this manuscript. Errors were found in the EHR data. Chart reviews suggested that maternal weight gain during pregnancy was measured with error in all records, with a median discrepancy of $-0.02$ kg/m$^2$ (range $-0.61$ to $0.93$). Childhood asthma was incorrectly recorded 10.4\% of times in the EHR (positive predictive value [PPV]= 0.57 and negative predictive value [NPV]=0.97). In contrast, childhood obesity was incorrect in only $0.6\%$ of records (PPV=0.998, NPV=0.991), although the time to obesity / censoring was incorrect in 4.7\% of reviewed records with a median discrepancy of 1.0 years (range 0.04, 1.8). Errors were found in all other variables, except for maternal age, in the EHR. Error rates ranged from rare (e.g., sex, 0.4\% misclassified) to fairly common (25\% with incorrect insurance status).

\subsection{Synthetic data for illustration}
For the purposes of open science we have created a simulated synthetic dataset based on the mother-child dataset of \citet{shepherd23}, so that others can download our code and data and run the examples presented in this manuscript. This synthetic data will be used throughout. The synthetic data share similar structures and error rates to the original data, with $N=10,335$ synthetic EHR mother-child dyads and a validation sample of $n=996$ mother-child dyads. Details for generating synthetic data are given in the Supplementary Materials S1. In this tutorial, we will consider the following outcomes: maternal weight gain during pregnancy (continuous), child asthma status (binary), and child time to obesity (time-to-event). We will consider fitting typical regression models (i.e., linear, logistic, and Cox proportional hazards) to estimate the association between covariates and these outcomes.

\section{Methods}
Several methods have been developed to address error in both $X$ and $Y$. These methods rely on data being available to inform the measurement error structure. We assume gold standard reference data (e.g. validation data) exist on a subset. In this section, we will show how to use this data to perform error-adjusted estimation and inference. The methods we consider include multiple imputation, sieve maximum likelihood estimation, inverse-probability weighting, and generalized raking. In the following subsections, we will describe each method and then discuss and illustrate its use with continuous, binary and time-to-event outcomes in the motivating MWG study. The variables involved in each regression are shown in Table 1. Code for the analyses discussed in this section are provided on GitHub (https://github.com/PamelaShaw/ErrorCorrectionReview).

\subsection{Multiple imputation} \label{mi}

Our setting, where we have error-prone data on all records but validated data on only a subset, can be thought of as a missing data problem. Specifically, we have complete data $(X^*,Y^*,Z,X,Y)$ on validated records ($R=1$), whereas we are missing $(X,Y)$ on those records that were not validated ($R=0$). Hence, statistical methods for addressing missing data may be applicable to address measurement error across multiple variables in the EHR when there is a validation sample. Ideally, investigators select records for validation in a probabilistic manner, not out of convenience. If so, and assuming that all records are available to be validated if selected, the missing mechanism is missing completely at random (MCAR; $(X,Y) \perp R$) if simple random sampling is used to select records for validation or missing at random (MAR; $(X,Y) \perp R | (X^*,Y^*,Z)$) if phase-1 data are used to select records for validation. In our mother-child obesity study, to improve statistical efficiency, phase-1 data were used to select records for validation; therefore, validation data were MAR. 

Multiple imputation is one of the most popular methods for addressing missing data, and it is also an excellent analysis choice for addressing errors in EHR data, particularly since the missingness mechanism for validation subsampling by design will be known to be MCAR or MAR. The basic idea is (i) to fit models for ($X,Y$) based on ($X^*,Y^*,Z$) among those in the validated dataset, (ii) to impute ($X,Y$) for those who were not in the validated dataset, (iii) to fit the model(s) of interest using the complete data (now available on everyone after imputation) and store the estimates of interest, (iv) to repeat steps (ii) and (iii) multiple times, and (iv) finally to combine estimates across imputation replications.  Multiple imputation has been used by many to address errors in a single variable (e.g., \citet{cole2006}; \citet{edwards13}), and it has also been used to address errors in multiple variables \citep{giganti20}. Multiple imputation can also be applied in a relatively straightforward manner to address otherwise challenging issues such as errors in inclusion criteria \citep{giganti20inclusion, austin23}.

Because of its flexibility and familiarity, multiple imputation is an excellent analysis choice to address error-prone EHR data -- but its use does have some potential drawbacks.  First, multiple imputation requires specifying imputation models, which are largely nuisance models, but correct specification of these imputation models is required to ensure unbiased estimation for parameters of interest. And with relatively small validation datasets, it may be difficult to fit accurate imputation models even though these models will be used to impute a lot of unvalidated data. Second, if imputation models are not compatible with analysis models, referred to as being uncongenial \citep{meng1994}, then MI variance estimators using the popular Rubin's rules will be biased. Uncongenial models can occur if the imputation model is more rich than the analysis model, which can happen when trying to fit realistic imputation models to error-prone data \citep{giganti20,giganti20inclusion}. This problem can be fixed by applying other methods for calculating variances \citep{RobinsWang2000} or bootstrapping \citep{bartlett2020}, but coding up these other methods can be labor intensive and bootstrapping multiply imputed data can be computationally intensive, depending on the size of the data.

\subsubsection{Multiple imputation illustration}

Here we illustrate the use of multiple imputation to correct measurement error using the synthetic mother-child weight study. Although there are many strategies for multiple imputation, we will illustrate using the popular \texttt{mice} package in \texttt{R}. 
 
There were no errors in maternal age at delivery. All other variables had at least some discrepancies between the EHR and what was discovered after chart review. Child sex was incorrectly recorded in only 3 of 996 records (0.3\%). In theory, one could impute validated child sex from phase-1 child sex; phase-1 child sex is simply extremely predictive of validated child sex. However, \texttt{mice} under default settings does not permit predictors with such high collinearity (it removes them and produces a warning); for simplicity, we assume true sex is the same as the sex given in the EHR for all records that were not validated. 

The linear regression analysis assessing factors associated with maternal weight change during pregnancy requires imputing validated data for 7 variables. The logistic regression analysis assessing factors associated with child asthma requires imputing validated data for 8 variables. The Cox regression analysis for factors associated with childhood obesity requires imputing 16 variables. For each of these analyses, we perform separate sets of multiple imputations. For the most part, we use defaults in \texttt{mice}: partial mean matching for continuous variables, logistic regression for binary variables, and polytomous logistic regression for the categorical variable race. With the childhood obesity analysis, phase-1 and phase-2 obesity are highly correlated (PPV and NPV of 0.995 and 0.998, respectively) as are phase-1 and phase-2 measures of the time-to-obesity/censoring (correlation 0.996). Because default \texttt{mice}, as mentioned above, cannot handle variables with correlation this strong, we imputed these variables by hand. Specifically, to impute childhood obesity status, we drew from the sampling distribution of the PPV and NPV of obesity; we then imputed validated data based on the sampled PPV, NPV and phase-1 obesity status. To impute time-to-obesity/censoring, we essentially sampled residuals from the validated data to the phase 1 data. Best practice for imputing covariates with time-to-event outcomes is to include the outcome indicator (obesity) and the baseline cumulative hazard as well as the other covariates and error-free covariates in the imputation model \citep{white2009}. Therefore, we computed the Nelson-Aalen cumulative hazard of obesity using the imputed obesity and the imputed time-to-obesity/censoring variables, and then used this cumulative hazard estimate with \texttt{mice} to impute the remaining validated values for the covariates. This process was repeated multiple times for our multiple imputation.

We present estimates using Rubin's rules for estimating the variance. Analyses were also repeated using the bootstrap, due to concerns that the imputation and analysis models were not congenial. We used the \texttt{R} package \texttt{bootImpute} to bootstrap (200 replications) and then impute (2 imputations within each bootstrap) for the linear and logistic regression analyses. Because the imputation procedure is slightly different between \texttt{mice} and \texttt{bootImpute}, point estimates are slightly different. For the Cox regression analysis, we coded the bootstrap ourselves (200 replications) and imputed as described above once within each bootstrap replication. 

Coefficient estimates and standard errors are given in Tables \ref{tab:weight-gain}, \ref{tab:asthma}, and \ref{tab:obesity}. Compared to the naive model, estimated coefficients using MI are generally in the same direction although the strength of association is often more pronounced (e.g., estimated coefficients for age, depression, public insurance, and smoking in Table \ref{tab:weight-gain}). Note that for the asthma and childhood obesity analyses (Tables \ref{tab:asthma} and \ref{tab:obesity}), phase-1 data were not available for some of variables so differences with naive estimates could be due to the non-collapsibility of estimates with logistic and Cox models, as well as errors in variables. Standard errors for MI estimates, which acknowledge the uncertainty in the accuracy of phase-1 data, are naturally larger than those for naive estimates. In the Supplementary Material (Tables \ref{tab:weight-gain-boot}, \ref{tab:asthma-boot}), we present estimates of coefficient standard errors based on the bootstrap. (We note that bootstrap confidence intervals are typically better estimated using percentiles or bias-correction techniques, than simple $\pm 1.96 \times$ standard error procedures \citep{efron93}.) In short, results were fairly similar using the bootstrap with \texttt{bootImpute} versus Rubin's rules with \texttt{mice} for the linear and logistic regression analyses, possibly because model uncongeniality was not very high in our setting. (Severe uncongeniality was seen in earlier work where multiple rows per subject were used in the imputation, which were then collapsed to a single row per subject for the analysis \citep{giganti20,giganti20inclusion}.) For the Cox regression analysis, bootstrap standard errors (presented in Table \ref{tab:obesity}) were substantially larger than those using Rubin's rules, as were 95\% confidence intervals.

\subsection{Maximum likelihood estimation/Sieve maximum likelihood}

An alternative analysis approach is maximum likelihood estimation. Consider the setting with continuous $Y$, where $Y=\beta_0 + \beta_1 X + \beta_2 Z + \epsilon$, $Y^*=Y+W$, $X^*=X+U$, and $W$ is potentially correlated with $(X,Z,U)$ such that $Y$ and $X$ are subject to additive, possibly dependent, errors. The goal is to estimate the parameters $\beta=(\beta_0, \beta_1, \beta_2)$. An observation, $(Y^*,X^*,W,U,Z)$ is assumed to be generated from the joint density

\begin{align*}
    p(Y^*,X^*,W,U,Z)&=p(Y^*|W,U,X^*,Z)p(W,U|X^*,Z)p(X^*,Z) \\
                &=p_{\beta}(Y|X,Z)p(W,U|X^*,Z)p(X^*,Z),
\end{align*} where the second line assumes that $\epsilon$ is independent of $W$ and $U$. With complete data, and models for $p_{\beta}(Y|X,Z)$ and $p(W,U|X^*,Z)$, MLEs can be derived by maximizing the log-likelihood,
$$\sum_{i=1}^N \log p(Y^*_i,X^*_i,W_i,U_i,Z_i) \propto \sum_{i=1}^N \left\{ \log p_{\beta}(Y_i|X_i,Z_i) + \log p(W_i,U_i|X^*_i,Z_i)\right\}.$$ (Note that $p(X^*,Z)$ can be estimated using the empirical distribution and factors out.) However, $W_i$ and $U_i$ are only known for the subset with $R_i=1$. Therefore, the observed data log-likelihood is of the form
\begin{align*}
    \sum_{i=1}^N R_i & \left\{\log p_{\beta}(Y_i|X_i,Z_i)  + \log p(W_i,U_i|X^*_i,Z_i)\right\} + \hspace{1 in} \\
& \sum_{i=1}^N (1-R_i) \log \int p_{\beta}(Y_i^*-w|X_i^*-u,Z_i) p(w,u|X_i^* ,Z_i)dw du.
\end{align*}
Maximization of this observed data log-likelihood is more computationally challenging because of the integration over unobserved values of $W$ and $U$ for those not validated, but it is still feasible. 

Consistent estimation of $\beta$ requires correct specification of both $p_{\beta}(Y|X,Z)$ and $p(W,U|X^*,Z)$. However, the latter model is a nuisance model. More flexible estimation would be to nonparametrically estimate $p(W,U|X^*,Z)$, i.e., estimation via an empirical distribution at each level of $X^*,Z$. This nonparametric estimation, however, is not feasible because of the curse of dimensionality if $(X^*,Z)$ are continuous or include many levels. An approach to address this difficulty is to use sieve maximum likelihood estimation (SMLE), where $p(w,u|X^*_i,Z_i)$ and $\log p(w,u|X^*_i,Z_i)$ are approximated using B-splines \citep{tao2021}.

Although written above for a continuous outcome, a similar estimation approach has been described for dichotomous outcomes, both using maximum likelihood estimation \citep{lyles2011} and sieve maximum likelihood estimation \citep{lotspeich2022}. 

Properly specified maximum likelihood estimators are the most efficient estimators, which is a major strength of these approaches. And very little efficiency is lost by using SMLE to flexibly estimate $p(W,U|X^*,Z)$. However, the B-spline approximations add complexity to the estimation procedure, and they have challenges if the dimensions of $(X^*,Z)$ are more than 2 or 3; further simplification / assumptions are required with larger numbers of covariates. In addition, despite their robustness to the distribution of $p(W,U|X^*,Z)$, SMLEs still require correct specification of the distribution of $p_{\beta}(Y|X,Z)$. This latter assumption is stronger than that made by some of the other analysis approaches such as the design-based estimators described below; design-based estimators specify a mean model (i.e., conditional expectation), but they do not assume an actual distribution (e.g., normality) for the outcome. Finally, the efficiency of MLEs/SMLEs comes with a certain lack of flexibility. Unlike MI and design-based approaches, which are generally available and easily programmed for a wide range of models, MLEs / SMLEs require substantial development / coding for each variation in the outcome type / assumptions. For example, methods and code for SMLEs with error-prone time-to-event data have not yet been developed, and their development is not necessarily a simple extension of these estimators for continuous or binary outcomes.

\subsubsection{SMLE Illustration}

The \texttt{sleev} package in R permits fitting SMLE with continuous or binary outcomes subject to outcome and covariate errors \citep{xiong2025}. We focus our illustration using this package (version 1.2.0). However, as highlighted above, SMLEs have challenges when there are many $(X^*,Z)$ variables, because estimating $p(W,U|X^*,Z)$ using B-splines is difficult due to the curse of dimensionality. In our EHR mother-child application, we have many $(X^*,Z)$. In particular, almost all of our variables are error-prone (i.e., can be regarded as $X^*$) and a quick look at the pairwise correlations between the errors $(W,U)$ and $(X^*,Z)$  (properly weighted to account for the biased sampling design) suggests that nearly all of $(X^*,Z)$ are associated with errors in at least one other variable. The \texttt{sleev} package, and SMLEs more generally, are unable to handle this level of complexity without some simplifications. One potential simplification that has been suggested is to collapse $(X^*,Z)$ into a smaller number of variables using principal components analysis or factor analysis of mixed data and then to fit models of $p(W,U|X^*,Z)$ with B-splines using these collapsed variables. This approach likely over-simplifies the relationship between the phase-1 data and errors, but simplifications such as this are necessary for non-parametric estimation. We use this approach in our illustration. %, recognizing that in some sense, we are stretching SMLE techniques to the edge (and perhaps beyond) of what they have been designed to handle. %However, we were unable to get our models using \texttt{sleev} to converge based on this strategy. 
Other options could be to fit a fully parametric model or to consider extensions of alternative semiparametric estimators to handle outcome errors (e.g., \citet{zhong2023}). Both options would require a substantial amount of coding; the latter would also require additional methods development. %Therefore, we will illustrate SMLE using a simplified regression model.

Consider estimating the relationship between maternal weight gain during pregnancy ($Y$) and the maternal characteristics $(X,Z)$ given in Table \ref{tab:models}. In addition to the intercept, this linear regression model requires estimating 9 coefficients (2 continuous variables, 4 dichotomous, and 1 categorical corresponding to 3 dummy variables; note that \texttt{sleev} requires coding dichotomous and categorical variables as dummy [0/1] variables). We use the \texttt{FAMD} function in the \texttt{FactoMineR} library to collapse these variables into factors. We then use the first two factors to flexibly estimate $p(W,U|X^*,Z)$. Note that we are essentially assuming that $p(W,U|X^*,Z)=p(W,U|A_1,A_2)$, where $(A_1,A_2,\cdots,A_9)$ is the full factorization of $(X^*,Z)$ and $(A_1,A_2)$ explains approximately 30\% of the variance. To fit the model of $p(W,U|A_1,A_2)$, we compute B-splines with degree=3 (corresponding to cubic splines, which are often favored in practice due to their smoothness) and size=4 (corresponding to the size of each of the two one-dimensional B-spline bases). When there are two continuous factors, $(A_1,A_2)$, the B-spline basis is constructed from the tensor product of the one-dimensional basis for each variable. Hence, the size of our multi-dimensional B-spline basis is 16 ($=4^2$). The degree and the size of the basis can be selected with cross-validation \citep{xiong2025}, but fitting a single model with our dataset takes multiple hours of computation time, so we have not considered alternative bases. To estimate the standard error of our estimated regression coefficients, we also have to be careful specifying \texttt{hn\_scale}, which controls the step size for variance estimation based on the profile likelihood \citep{murphy2000}. We chose \texttt{hn\_scale}$=1/100$ corresponding to step sizes of $1/(100\sqrt{N}) \approx 10^{-4}$ because of the small values of the regression coefficients. Variance estimates were nearly identical with \texttt{hn\_scale}$=1/200$ (which suggests a good choice of step size) but did not converge with \texttt{hn\_scale}$=1/2$ or 1 (the default). Coefficient estimates and standard errors are given in Table \ref{tab:weight-gain}. Estimates are generally somewhat similar to those based on the other approaches with the exception of the regression coefficient for depression, public insurance, and tobacco use, which interestingly happen to be some of our most error-prone covariates. In addition, standard errors for the SMLE estimates tend to be smaller than those of the other methods (not surprising) and of around the same size as the naive estimates (a bit more surprising).

For the binary outcome of childhood asthma, we followed a similar strategy by first collapsing $(X^*,Z)$ using the \texttt{FAMD} function and taking the first two components, fitting a multi-dimensional B-spline basis of size$=16$ and degree$=3$, and then fitting the SMLE with \texttt{hn\_scale}$=0.1$. Coefficient estimates and standard errors are given in Table \ref{tab:asthma}. For the most part, estimates are roughly similar to those of the other estimation techniques. Standard errors tend to be smaller than most of the other two-phase analysis approaches (most notably for the weight change and EGA coefficients) but not always (e.g., public insurance and black race coefficients).

%we were unable to obtain SMLEs using \texttt{sleev} for our desired model following a similar strategy to that described above for the continuous outcome. If we further simplified our model by removing estimated gestational age, assuming that race was measured in the phase-1 data without errors, and summarizing $(X^*,Z)$ to their first two factors, then we were able to obtain point estimates for logistic regression coefficients. However, we were unable to estimate standard errors in this model. 
%Again, this analysis stretches the limits of the SMLE implementation and required several simplifying assumptions; hence, we are nervous about the use of SMLE in this setting. 

We were unable to obtain SMLEs for our time-to-event analysis, because SMLEs have not yet been developed to handle this setting.

%compute B-splines. maternal BMI before pregnancy ($X$) adjusting for maternal age ($Z$). Because $X^*$ and $Z$ are both continuous, we still need to do some data collapsing to use B-splines to estimate $p(W,U|X^*,Z)$. We use principal components analysis and construct B-splines using the first principal component of $(X^*,Z)$. The B-splines had size 20 and order=3; results were similar with B-splines of size 30. Then these B-splines were used for the estimation of $P(W,U|X^*,Z)$, and finally SMLEs were computed by maximizing the semiparametric likelihood. Analysis code is provided in the Supplemental Material.

%The resulting SMLEs and standard errors for the model of $E(Y|X,Z)=\beta_0+\beta_XX + \beta_ZZ$ are $\hat \beta = (\hat \beta_0, \hat \beta_X, \hat \beta_Z)=(0.4231, -0.0036, -0.0010)$ and $(3.65,0.34,0.33) \times 10^{-3}$, respectively. For comparison, the naive linear regression analysis using error-prone phase-1 data produced estimates and standard errors of $(0.4613, -0.0044, -0.0011)$ and $(12.9,0.33,0.38) \times 10^{-3}$, respectively. 

\subsection{Design-based methods} \label{design-based}

As seen in section \ref{mi}, traditional missing data approaches can be successful at addressing measurement error.  In design-based approaches, one models the probability of having complete data but avoids making any modeling assumptions about the outcome in the individuals with missing (i.e., error-prone) data. In this section, we will consider two popular and easy to implement, design-based approaches for addressing missing data: inverse probability weighting (IPW) (\cite{horvitz1952generalization,seaman2013review}) and generalized raking (\cite{deville1993generalized,deville92}), which is equivalent to  a type of augmented inverse probablity weighting (AIPW) \citep{lumley11}. Modeling the missingness would generally be a more straightforward task in the setting of probabilistic validation subsampling, where the missingness mechansim would be known.  Estimators are then weighted equations of the observed validated data. Thus, design-based estimators are generally more robust than multiple imputation approaches that leverage efficiency from making modeling assumptions for the outcome in the unobserved data. While inverse-weighted estimators are often looked upon unfavorably due to expected efficiency loss,  \citep{lumley17} shows that even in the setting where is slight (undetectable) model mis-specification, the excess variability in an efficient augmented inverse probability weighted (AIPW) is comparable to the potential bias in MI from model misclassification. Design-based estimators are applicable under very mild regularity assumptions, which generally relate to well-behaved distributions for the data and the probability of being observed being positive for all individuals \citep{sarndal2003model}. 
%%When stratified random sampling is used, such is the case with our synthetic maternal weight data set, one can use the observed sampling fractions in the strata for $\hat{\pi}_i$, commonly referred to as post-stratification.  
Applying IPW in the measurement error setting is no different than when it is applied in the missing data setting. In our setting the missing data is the gold standard data (X,Y) on the subset not in the validation subset. One either knows or needs to build a model the probability of an individual having complete data observed, conditional on an individual's characteristics. That is, one builds a model for $\pi_i=P(R_i=1|V_i)$, where $V=(X^*,Y^*,Z)$ i.e. the data available on the whole cohort. The inverse probability weights are then $1/{\hat{\pi}_i}$. These weights are also known as the Horwitz-Thompson weights \citep{horvitz1952generalization}. Commonly, a parametric logistic regression is used to model $\pi_i$. Even in case where the validation subset was selected by design, IPW is expected to be more efficient to use estimated weights instead of the known sampling weights \citep{robins94,lumley11}. One estimates the parameter $\beta$ by fitting a weighed estimating equation of the complete (validated) data $\Sigma_{i=1}^N \frac{R_i}{\hat{\pi}_i} \mathcal{P}(\beta;X,Z,Y)=0$, where $\mathcal{P}(\beta;X,Z,Y)$ is the estimating equation for the parameters in the target outcome model.  To account for the uncertainty in weights estimated by a model, one can use either the stacked estimating equation sandwich approach \citep{stefanski2002calculus} or the bootstrap \citep{efron93} for correct inference. \cite{little2024comparison} provide a general guidance for the application of IPW, including selection of variables for the missingness model and settings where you would expect it to do be more efficient than multiple imputation. Generally speaking, when there is some misspecification in the imputation model, IPW can be an improvement over MI with respect to multiple imputation; however, because IPW is not using all the data in the outcome regression, IPW can be highly inefficient relative to an approach like MI that does leverage this information.
%that incorporates inforamtion from all indivdiuals into the analysis through an adjustgment (calibration) of the expected to be an improvement over IPW whenever there are auxiliary data available in the cohort of individuals with missing data that can leverage information in the individuals with only partially observed data. Given the expense of validation, the proportion of individuals who are missing the error-free data could be quite large, and in many settings there could be a lot of additional data observed on those individuals.
Generalized raking (GR), also known as survey calibration, is a type of AIPW estimator \citep{lumley11,robins94}. For the GR approach, one uses auxiliary data to adjust the usual Horwitz-Thompson weights in a way that leverages the information in the non-validated individuals regarding the target parameters. Specifically, one finds a multiplicative adjustment factor for the weights, $g_i$,  such that it minimizes the distance to the Horwitz-Thompson weights while satisfying a constraint defined by the auxiliary variables. Namely, for a given distance function d(a,b) one finds $g_i$ that
\begin{eqnarray*}
& \mbox{minimize } \Sigma_{i=1}^N d(\frac{g_i}{\pi_i},\frac{1}{\pi_i}) \\
& \mbox{subject to } \Sigma_{i=1}^N R_i\frac{g_i}{\pi_i} A_i =  \Sigma_{i=1}^N A_i.
\end{eqnarray*}
\noindent Commonly,d(a,b) is chosen to be the Poisson deviance $a \log(a/b) + (b-a)$, which will yields non-negative weights  \cite{breslow09} showed that when the auxiliary variable $A_i$ is the expected value of the influence function for the target parameter given the observed data, this approach will be the maximally efficient design-based estimator and that whenever the correlation between $A_i$ and this quantity is positive, the GR estimator will be more efficient than IPW.  \cite{breslow09} also demonstrated how to practically apply this approach following the suggested procedure by \cite{kulich04}, which is to fit the targeted estimating equations using the error prone data first and use the influence functions for the parameters from this model as the set of auxiliary variables to calibrate the Horwitz-Thompson weights. As will be discussed in the illustration, below, this approach is straightforward to obtain using standard functions in R for many models. \cite{han21} and \cite{oh21} demonstrated there can be efficiency gains by using multiple imputation to multiple impute the error free data or to directly model the missing influence function rather than the error-free data. Though in other practical settings, the simpler approach of using the so-called naive influence functions from the error-prone model which have been observed to be nearly as efficient in practical settings \citep{shepherd23}.

There have been a few studies to date that have used this approach to analyze data with errors in both the outcome and exposures. For error-prone covariates and survival outcomes, \cite{oh20} compared the application of generalized raking to regression calibration, showing it generally had superior performance with little to no bias and good efficiency relative to IPW. These authors also found using multiple imputation for the auxiliary variable or by modeling potential non-linear relationships between the observed data and target influence functions led to some gains in efficiency for the raking estimator compared to the \cite{kulich04} approach using the naive influence functions from the model fit to the error-prone data \citep{oh21}.  \cite{shepherd23} applied these methods in a setting where the validation subset was selected using an optimal stratified sampling scheme and found some efficiency gain in the target parameter (maternal weight gain) for raking relative to IPW for the child obesity outcome, but no gain for the asthma outcome. This was somewhat of an unfair comparison because the validation design optimal for IPW is not the same as the optimal design for raking \citep{chen2022optimal,yang2025optimal}; however, one should also expect for an efficient validation design that the choice of estimators will have less of an impact on the efficiency of the final estimates. 

\subsubsection{Design-based methods illustration} \label{design-based-i}

A first step for both the IPW and GR analyses is to estimate the probability that a particular observation has complete data (i.e., is selected into the phase-2 subset) given the observed phase-1 data. In the maternal-child study, the phase-2 subset was chosen by design by stratified sampling, and thus the probability of being sampled is $n_s/N_s$, where $n_s$ and $N_s$ are the stratum sizes for stratum $s$ in the phase 2 subset and phase 1 population, respectively. Another common setting is that the probability of selection needs to be modeled, commonly by a logistic regression model for the selection indicator (R=1) using variables that are thought to be plausibly related to both the missingness mechanism and the outcome \citep{little2024comparison,seaman2013review}
%\[ logit (R_i) = \alpha_0 + \alpha_1 BMIChangeEst + \alpha_2 Age + \alpha_3 Asian + \alpha_4 Black + %\alpha_5 Other+ \alpha_6 Hispanic + \alpha_7 Depression + \alpha_8 Public Insurance + \alpha_9 %Smoking.\]

The Horwtiz-Thompson weight for observation $i$ is the inverted predicted probability of being selected $1/\pi_i$. The IPW analysis can be done using the \texttt{survey} package for a variety of two-phase designs, including the missing not be design case (\cite{lumleyBook}. In the case where weights must be estimated, the estWeights function can be used which will use a sandwich variance estimate to account for the uncertainty of the weights. We provide code in the appendix for our data example which used stratified sampled, but sample R code for the estimated weights example is available in this online vignette (https://kpwhri.github.io/actstats/IPWDemo20260112.html). Table 2-4  shows the results.  The SE for most variables  for the continuous and binary outcomes was lower for GR compared to IPW, with reductions up to approximately 50\% for the continuous outcome and 20\%  for the binary outcome. For Cox regression, results were mixed. GR reduced the SE for a few of the coefficients but others were little changed or even increased.   This is likely due to there being too many constraints given there are only 413 observed events in the validated data. The constraints include the 18 estimating equations for the regression parameters plus 51 additional calibration constraints (18 for the influence functions for each regression coefficient plus the 33 for the sampling strata). Applying the GR procedure calibrating only with the 33 sampling strata and one influence function for the variable of interest, weight change, lowered the SE slightly for the raking estimator (data not shown), suggesting too many constraints is contributing to variance inflation.

\subsection{Other methods}

There are other common methods for addressing covariate measurement error including regression calibration, SIMEX, and general Bayesian methods. To our knowledge, little to now work has been done to extend these methods to address measurement error in both outcomes and covariates. Here we briefly touch on these other approaches and consider challenges.

\subsubsection{Regression calibration}

%Moment-based approach. Work has been done with continuous $X$ and continuous $Y$ (Shepherd and Yu; our papers together). Boutique examples. Etc.

% add sentence aobut Moment reconstruction 

Regression calibration (RC) is a method introduced by \cite{prentice82} to address covariate measurement error when performing a regression of time-to-event outcomes and has since become one of the most widely used methods to address covariate measurement error \citep{carroll06}. This method is an intuitive approach where, in the regression of interest, the unobserved exposure $X$ is replaced by an estimate $\hat{X}=E[X|X^\star,Z,V]$, where $Z$ is the vector of other covariates in the outcome regression and $V$ may be other precision variables that are unrelated to the outcome, but help the precision of $\hat{X}$. To apply RC, one only needs to have either the true $X$ in a subset or a second measure $X^{\star\star}$ with classical measurement error that is independent of the error in $X^\star$. One then regresses $X^{\star \star}$ on the covariate vector $(X^*,Z,V)$ and generates the estimated value $\hat{X}$ from the fitted regression. Its popularity stems from ease of application and from its good performance in several settings \cite{carroll06}.  In particular, when $\hat{X}$ is correctly modeled for a continuous $X$, this approach will be consistent in linear models \cite{grace2021handbook}. In non-linear models, regression calibration will typically have some bias; however, RC can be seen as a first order (linear) approximation, which is very stable and can have lower mean-squared error than more sophisticated consistent approaches if the problem is sufficiently linear. In Cox and logistic regression, RC has been seen to out perform the corrected sore and conditional score approaches to address covariate measurement error when the association with $X$ and measurement error variance is moderate and the event rate is low, and  \cite{sugar2007logistic, shaw12}. 

Regression calibration has been used less often to adjust for error in the outcome. Simply replacing a continuous $Y$ with a predicted value in the target regression will not generally be a good strategy.  In the linear model, Berkson error in a predicted outcome will induce bias in the estimated regression coefficients \cite{keogh20}. There are specialized settings in which RC has been used successfully to adjust for error in the outcome and covariate. \cite{shepherd12} applied regression calibration in the linear outcome model setting where there were errors in both $X$ and $Y$, which were potentially correlated and which followed a linear measurement error. The adjustment for the error in $X$ was essentially the traditional application of RC, whereas the adjustment for the error in $Y$ was done in a way that subtracted off an estimated bias term. \cite{boe20} applied RC to address measurement error in the covariate in the Cox model, while also simultaneously addressing misclassification in a survival outcome. \cite{oh20} applied RC to adjust for error only in a survival outcome in a setting where there was validation data on a subset, and this RC thus has potential to be used in the setting where $X$ has errors addressable by RC. This approach has the same weakness RC has more generally. \cite{oh20} found their RC approach only performed well in terms of lower MSE, relative to GR (Section \ref{design-based} when there was no misclassification in the event indicator, low-to-moderate error, and high censoring. 

Regression calibration has been applied to settings with multiple error-prone continuous covariates, but is not expected to do well in highly non-linear settings due to it being only a first order linear approximation. While some adjustments have been made to improve its performance for non-linear settings \citep{carroll06}, applying these methods in multivariate settings or settings with both covariate and outcome error, have not studied. 

\subsubsection{SIMEX}
The Simulation Extroplation (SIMEX) method is another popular method that was introduced for the linear regression setting where there is classical (mean zero random additive) error with known error variance in a single continuous exposure \citep{cook94}. The idea is that additional error is added to the error-prone covariate and then from the resulting range of estimates, a curve as a function of the amount of error variance is fit and the value of beta when there is no error is then extrapolated from this curve. The success of SIMEX as a measurement error correction method relies on the correct specification of the extrapolant curve. For the simple linear regression problem, the form for the extrapolant function is known, but for most settings it would likely need to be estimated \citep{cook94}. SIMEX has become popular due to its intuitive approach and has been extended to a variety of settings \citep{sevilimedu2022simulation}, but generally there has been very limited work applying SIMEX to address error in multiple variables. Two complications are that multivariate error distributions would need to be generated in the Simulation step and the form of the extrapolant function for the extrapolation step, and it is not clear to what degree the covariance structure of the multivariate measurement error would need to be incorporated. The extrapolant function, a key to SIMEX's success, would likely depend on the setting and something difficult to derive and fit. In our Maternal Weight gain example, there are up to 18 error-prone covariates with an additionally error-prone outcome, and thus no SIMEX algorithm or software is readily available for this setting. 

%in the regression and deriving the form of such an extrapolant function given both variables and outcomes could be non-li %For example, when covariates are correlated, error in one covariate will cause bias in the coefficient of the other regression parameters of the other covariates.  Thus, the extrapolation and simulation step would likely need to be multivariate in nature without making very strong, likely unverifiable assumptions.

%WHat I just wrote above:  
%SIMEX has been seen to work very well with covariate meeasuremente error. With errors in outcome and exposures, imagine simulating with varying levels of error in two variables and then extrapolating in the 2-dimensional setting back to 0 to get estimates. It is unclear how well this approach would work with higher dimensional measurement error with correlation across variables. It could become unweildy -- in our setting we have up to 16 error-prone variables that would all require simulating and extracting. There are also questions with how to treat the dependence parameters across variables; natural choice would be to assume the same amount seen in the phase-2 data, but this may also be something that needs to be considered.

\subsubsection{Bayesian methods}
Bayesian methods to address measurement error are quite flexible, but do require that models be specified for the structure of the errors in the variables, in addition to the usual modeling components of a Bayesian analysis approach \citep{gustafson03}. This may be difficult to do in our setting, unless there was strong prior knowledge about this structure given validation data often make up a relatively small proportion of the data. In our problem, we would need to make distributional assumptions about the structure of the outcome and 18 covariates, as well as how they relate to all the errorprone versions. In our motivating setting, we had very little prior knowledge of the structure of the measurement error.  Additionally in some settings, such as those with large EHR datasets,  computational challenges could arise.  

\subsubsection{Quantitative bias analysis}
In the absence of any validation data to inform the measurement error structure, one could consider the framework of quantitative bias analysis as a way to conduct a sensitivity analysis under different assumptions for the measurement error in the study variables \citep{petersen2021systematic,lash2014good,greenland1996basic}. \cite{shaw20} provide a general overview of considerations for this method in the context of measurement error and provide examples of authors who have applied this approach for the setting where a single exposure was prone to measurement error. Detailed frameworks for settings such as ours, where 20 or more variables have potential measurement error have not appeared in the literature to date.

\section{Discussion}

In this manuscript, we have described approaches for addressing errors across multiple variables when validation data are available. With widespread use of routinely collected data for research, such methods are important. 

A few of the methods we have presented, MI, IPW, and generalized raking, can be readily implemented in a wide variety of analysis settings. In contrast, SMLE, is not yet as widely applicable, and warrants additional development. Other methods based on regression calibration and SIMEX are much less developed, and it is unclear whether additional developments are warranted given inherent challenges.

In our examples, we saw that SMLE was generally more efficient (lower standard errors) than the other approaches, followed by MI, GR, and IPW. As SMLE is a likelihood-based approach, this result is not surprising. However, SMLE lacks the flexibility of the other three approaches and current implementation requires strong modeling assumptions and/or the collapsing of multiple variables into smaller numbers of factors. 

MI is readily applicable in a wide variety of settings and is an excellent analysis option. However, its limitations should be recognized, particularly the requirement of selecting an imputation model. MIs gains in efficiency over GR come with losses in robustness.

IPW and GR are the most robust approaches. They often have slightly less efficiency than MI, but not always. GR is asymptotically guaranteed to be no worse than MI, so we believe it always is preferable to IPW. Often the efficiency gains of GR over IPW are substantial, as seen in the maternal weight gain and asthma examples.  In these cases the calibration variables had good linear correlation with the influence fucntions of the target influence functions for the paramters of interst. However, as seen in the time to obesity example, sometimes the benefits of raking are minimal and in limited samples the calibration constraints could cause some inflation in the variance relative to IPW.  The gains in efficiency for GR have to do with the adequacy of the auxiliary variable as well as sample size. With error-prone data, it is often easy to construct auxiliary variables which are quite good- such as in our EHR setting where the measurement error in many variables was modest. The GR approach can sometimes be improved by by multiply imputing the auxiliary variable (\cite{han21}), the estimated coefficient influence functions, but in our setting, we did not see much gain in multiply imputing the auxiliary variable over just using naive estimates of it.

In any given setting, the complexity of the analysis model and expert knowledge of the potential structure of the measurement error in analysis variables can help guide the choice of method. Currently multiple imputation and generalized raking are two easy to implement methods that can be applied to a wide variety of settings. Both these approaches do depend at least in part on modeling assumptions regarding the structure of the data. More flexible semi-parametric and non-parametric approaches that make fewer assumptions and therefore offer more robustness are important areas of future research. For these methods to be truly useful to practitioners, they would need to be readily scalable to large datasets and user friendly sofware would to implement these methods would also be necessary.

\clearpage
%%% Need to update the bibliographic style once get into Wiley template
\bibliographystyle{biom}
\bibliography{SIM_Method_Tutorial}
\clearpage
\newpage

\begin{table}[ht]

\caption{Outcome models considered for illustration}
\vspace{.1 in}
\begin{tabular}{|l| l | l |}
\hline
Outcome model & Outcome & Covariates  \\ 
  \hline
  Linear & Pregnancy weight & body mass index, age, race (4-level), Hispanic ethnicity,\\
   & change &   depression, public insurance, smoking  \\
  & & \\
Logistic & Asthma & Pregnancy weight change, body mass index, Age, race \\
 & &   (3-level), male sex, public insurance, smoking, maternal  \\
  & &   asthma, gestational age \\

   & & \\
Cox  & Time to obesity & Pregnancy weight change, body mass index, age, race    \\
proportional& & (4-level), Hispanic ethnicity, gestational diabetes, type I/II \\
hazards & &    diabetes, Cesarean, male sex, depression, public insurance,  \\
   & & singleton, smoking, married, prior children number, \\
      & &  gestational age\\
      & & \\
\hline
\end{tabular}
\label{tab:models}
\end{table}
\clearpage
\newpage

\begin{table}[ht]
\caption{Regression coefficients (standard errors) for weight gain during pregnancy using the various analysis approaches.}

\begin{tabular}{rrrrrr}
  \hline
Coefficient & Naive & IPW & Raking & MI & SMLE \\ 
  \hline
Intercept & 0.487 (0.014) & 0.530 (0.045) & 0.514 (0.022) & 0.503 (0.021) & 0.466 (0.013)\\ 
  BMI$^*$ & -0.022 (0.002) & -0.014 (0.005) & -0.018 (0.003) & -0.015 (0.003) & -0.019 (0.002)\\ 
  Age$^*$ & -0.016 (0.004) & -0.040 (0.014) & -0.030 (0.007) & -0.032 (0.006) & -0.016 (0.004)\\ 
  Asian & -0.039 (0.010) & -0.008 (0.040) & -0.039 (0.017) & -0.036 (0.016) & -0.041 (0.009)\\ 
  Black & -0.019 (0.005) & -0.012 (0.017) & -0.021 (0.009) & -0.032 (0.007) & -0.021 (0.005)\\ 
  Other & -0.011 (0.008) & -0.018 (0.026) & -0.024 (0.023) & -0.020 (0.010) & -0.009 (0.008)\\ 
  Hispanic & 0.007 (0.007) & 0.032 (0.024) & 0.022 (0.015) & 0.006 (0.013) & 0.006 (0.007)\\ 
  Depression & 0.008 (0.007) & 0.065 (0.026) & 0.065 (0.025) & 0.038 (0.024) & -0.004 (0.006)\\ 
  Public Insurance & -0.017 (0.005) & -0.064 (0.017) & -0.052 (0.015) & -0.030 (0.020) & -0.010 (0.004)\\ 
  Smoking & -0.011 (0.010) & -0.024 (0.022) & -0.030 (0.020) & -0.032 (0.020) & -0.005 (0.008)\\ 
   \hline
\end{tabular}
\footnotesize{* Body Mass Index (BMI) coefficient is per 5 kg$/m^2$; Age coefficient is per 10 years}
\label{tab:weight-gain}
\end{table}
\clearpage
\newpage

\begin{table}[ht]
\caption{Regression coefficients (standard errors) for logistic regression model of childhood asthma using the various analysis approaches.} %$^*$To get the SMLE model to fit, we removed EGA and assumed that race was error-free; even after these simplifications, we were unable to obtain standard error estimates.}

\begin{tabular}{rrrrrr}
  \hline
Coefficient & Naive & IPW & Raking & MI & SMLE \\ 
  \hline
Intercept & -2.441 (0.210) & -4.351 (2.326) &-4.109 (2.175) & -1.846 (2.072) & -2.330 (0.176) \\ 
Weight Change & 0.205 (0.142) & 1.013 (0.815) & 1.570 (0.722) & 1.153 (0.459) & 1.134 (0.092)\\
  BMI$^*$ & 0.128 (0.023) & 0.259 (0.099) & 0.255 (0.087) & 0.173 (0.051) & 0.197 (0.064)\\ 
  Age$^*$ & -0.198 (0.056) & -0.465 (0.222) & -0.278 (0.201) & -0.271 (0.163) & -0.342 (0.150)\\ 
  Black & 0.936 (0.069) & 1.333 (0.332) & 1.269 (0.277) & 1.100 (0.165) & 1.224 (0.206)\\ 
  Other & 0.228 (0.098) & 1.061 (0.394) & 0.983 (0.387) & 0.655 (0.270) & 1.054 (0.226)\\ 
  Male Sex & 0.553 (0.064) & 1.087 (0.301) & 0.899 (0.240) & 0.764 (0.195) & 0.764 (0.183)\\
  Public Insurance & 0.183 (0.063) & 0.379 (0.353) & 0.403 (0.342) & 0.358 (0.268) &  0.491 (0.339)\\ 
  Smoking & 0.056 (0.137) & 0.773 (0.384) & 0.823 (0.385) & 0.373 (0.344) & 0.470 (0.293)\\ 
  Maternal Asthma & 0.638 (0.099) & -0.238 (0.396)& 0.171 (0.380) & 0.578 (0.294) & 0.455 (0.308) \\ 
  EGA$^*$ &  & 0.006 (0.057) & -0.013 (0.055) & -0.046 (0.046) & -0.041 (0.014)\\
   \hline
\end{tabular}
\footnotesize{* Body Mass Index (BMI) coefficient is per 5 kg$/m^2$; Age coefficient is per 10 years; EGA: estimated gestational age}
\label{tab:asthma}
\end{table}
\clearpage
\newpage

\begin{table}[ht]
\caption{Regression coefficients (standard errors) for Cox regression model of child's time-to-obesity using the various analysis approaches.}
\begin{center}

\begin{tabular}{rrrrr}
  \hline
Coefficient & Naive & IPW & Raking & MI (RR) (Boot)  \\ 
  \hline
Weight Change & 1.043 (0.100) & 1.183 (0.388) & 1.048 (0.392) & 1.237 (0.168) (0.212)\\
  BMI$^*$ & 0.317 (0.016) & 0.273 (0.048) & 0.308 (0.049) & 0.326 (0.021) (0.026)\\ 
  Age$^*$ & -0.032 (0.040) &-0.074 (0.157) & 0.024 (0.157) & 0.043 (0.073) (0.123)\\ 
  Asian & 0.209 (0.105) & 0.277 (0.404) & 0.266 (0.410) & 0.236 (0.161) (0.299)\\
  Black & -0.179 (0.058) & -0.271 (0.210) &-0.092 (0.204) & -0.196 (0.089) (0.110)\\ 
  Other & 0.136 (0.073) & 0.142 (0.222) & 0.291 (0.222) & 0.253 (0.099) (0.194)\\ 
  Hispanic & 0.794 (0.055)& 0.847 (0.201) & 0.796 (0.204) & 0.566 (0.109) (0.191)\\
  Gestational Diabetes & -0.271 (0.065) & -0.442 (0.306) & -0.338 (0.293) & -0.586 (0.134) (0.239) \\
  Type I/II Diabetes & -0.168 (0.126) & 0.425 (0.396) & 0.290 (0.407) & -0.141 (0.158) (0.385)\\
  Cesarean & 0.213 (0.046) & 0.409 (0.154) & 0.220 (0.154) & 0.182 (0.087) (0.143)\\
  Male Sex & -0.109 (0.044) & 0.069 (0.147) & -0.143 (0.146) & -0.147 (0.048) (0.062)\\
  Depression & -0.107 (0.071) & 0.195 (0.242) & 0.133 (0.245) & -0.209 (0.200) (0.232)\\
  Public Insurance & 0.174 (0.046) & 0.466 (0.215) & 0.443 (0.212) & 0.368 (0.127) (0.218)\\ 
  Singleton & 0.606 (0.215) & 0.355 (0.745) & 0.188 (0.718) & 0.264 (0.281) (0.409)\\
  Smoking & 0.216 (0.090) & 0.561 (0.216) & 0.601 (0.213) & 0.352 (0.153) (0.226)\\ 
  Married & & 0.139 (0.199) & 0.162 (0.199) & 0.101 (0.158) (0.219)\\
  No. of prior children & & -0.067 (0.085) & -0.092 (0.084) & -0.064 (0.047) (0.094)\\
  EGA$^*$ &  & 0.024 (0.035) & 0.025 (0.035) & 0.025 (0.026) (0.037)\\
   \hline
\end{tabular}
\footnotesize{* Body Mass Index (BMI) coefficient is per 5 kg$/m^2$; Age coefficient is per 10 years; EGA: estimated gestational age}
\end{center}
\label{tab:obesity}
\end{table}

\clearpage
\large
\section*{Supplementary Materials}
\normalfont
\newpage

\subsection*{S1 Generating Synthetic Maternal Weight Gain (MWG) Study Data}

The R code used to generate the synthetic data is available at https://github.com/PamelaShaw/ErrorCorrectionReview. The general strategy for generating synthetic data was the following:

\begin{enumerate}
    \item Fit properly weighted models to the original MWG study data. Specifically, weighting based on the inverse sampling probabilities in the original study, we fit the following models / summaries of the original phase-2 (error-free) data $(X,Y)$:
    \begin{enumerate}
      \item Cross tabulation: race and ethnicity
      \item Logistic regression: Insurance status $\sim$ race + ethnicity
      \item Logistic regression: Tobacco use $\sim$ insurance + race + ethnicity
      \item Logistic regression: Singleton $\sim$ tobacco + insurance + race + ethnicity
      \item Logistic regression: Depression $\sim$ singleton + tobacco + insurance + race + ethnicity
      \item Logistic regression: Cesarean $\sim$ depression + singleton + tobacco + insurance + race + ethnicity
      \item Logistic regression: Sex(child) $\sim$ Cesarean + depression + singleton + tobacco + insurance + race + ethnicity
      \item Logistic regression: Married $\sim$ sex + Cesarean + depression + singleton + tobacco + insurance + race + ethnicity
      \item Logistic regression: Asthma $\sim$ married + sex + Cesarean + depression + singleton + tobacco + insurance + race + ethnicity
      \item Poisson regression: Number of children $\sim$ asthma + married + sex + Cesarean + depression + singleton + tobacco + insurance + race + ethnicity
      \item Multinomial logistic regression: Diabetes $\sim$ no.children + asthma + married + sex + Cesarean + depression + singleton + tobacco + insurance + race + ethnicity
      \item Linear regression: Age $\sim$ diabetes + no.children + asthma + married + sex + Cesarean + depression + singleton + tobacco + insurance + race + ethnicity
      \item Linear regression: log(45-EGA) $\sim$ age + diabetes + no.children + asthma + married + sex + Cesarean + depression + singleton + tobacco + insurance + race + ethnicity
      \item Linear regression: log(BMI) $\sim$ EGA + age + diabetes + no.children + asthma + married + sex + Cesarean + depression + singleton + tobacco + insurance + race + ethnicity
      \item Linear regression: weight change $\sim$ BMI + BMI$^2$ + EGA + age + diabetes + no.children + asthma + married + sex + Cesarean + depression + singleton + tobacco + insurance + race + ethnicity
      \item Weibull model: (time-to-obesity, obesity) $\sim$ weight change + BMI + age + diabetes + no.children + asthma + married + sex + Cesarean + depression + singleton + tobacco + insurance + race + ethnicity (weights calibrated with the naive influence function for weight change coefficient and sampling strata)
    \end{enumerate}
    \item Simulate synthetic error-free data $(X,Y)$ from these fitted models for $N=10,335$ records. These data will later be sampled for the synthetic phase-2 study data. Data were generated sequentially. We started by generating synthetic race and ethnicity using cross-tabulation (a). Then based on synthetic race and ethnicity, we simulated synthetic insurance status by Bernoulli draws from the predicted probability of having insurance based on model (b). Synthetic tobacco use was then simulated using fitted model (c) based on synthetic race, ethnicity, and insurance in a similar manner. This process continued sequentially, generating data from fitted models (d) through (p) using the already generated synthetic data as input variables. When generating synthetic continuous data, we drew from normal distributions with variance based on models' residual variance, and we put bounds on the values roughly based on the range seen in the original data. We generated synthetic time-to-obesity from model (p); then generated synthetic time-to-censoring from a mixture distribution of a uniform(2,6) distribution and a point mass at 6 with the mixing parameter (probability of being the point mass) equal to 0.485; and finally computed the synthetic observed time-to-obesity/censoring as the minimum of these two values and synthetic obesity as 1 if synthetic time-to-obesity was before synthetic time-to-censoring, otherwise 0.
    \item Fit models estimating the rates of errors and magnitude of errors by comparing the phase 2 (error-free) data $(X,Y)$ with the error-prone phase 1 data $(X^*,Y^*)$ in the original MWG study data. Then generate synthetic error-prone data $(X^*,Y^*)$ from these models for $N=10,335$ records. These are the synthetic phase-1 data. Specifically,
    \begin{enumerate}
      \item Synthetic phase-1 maternal age was set to equal error-free synthetic maternal age, as there were no errors in the original data for this variable.
      \item The differences between phase-1 and phase-2 BMI and weight change in the original data were calculated, creating pairs of errors. We then sampled with replacement from these paired errors and added them to the synthetic error-free values of these variables, along with some minor jittering (from normal distributions with means 0 and standard deviations 0.1 and 0.05, respectively).
      \item The true positive and true negative rates for obesity were calculated based on the original phase-1 and phase-2 data, and synthetic phase-1 data were then generated with Bernoulli draws from these estimated true positive and true negative rates given the synthetic error-free obesity status.
      \item The differences in the original data between the phase-1 and phase-2 times-to-obesity, conditional on whether there were errors in the obesity status, were computed. Synthetic time-to-obesity data were then generated by sampling with replacement from these errors, adding them to the synthetic error-free time-to-obesity values (conditional on whether or not there were errors in the synthetic obesity status), adding some jittering noise (normal distribution with mean 0 and standard deviation 0.1), and then truncating to ensure that all values were between 2.001 and 6.
      \item For the remaining binary variables, tobacco use, singleton status, depression, cesarean indicator, male sex, Hispanic ethnicity, insurance status, and asthma, the true positive and true negative rates were estimated from the original phase-1 and phase-2 data. To create synthetic phase-1 data for these variables, we then sampled from Bernoulli distributions conditional on the synthetic error-free data using these estimated rates for each variable.
      \item For the multinomial variables, maternal race and maternal diabetes status, we similarly estimated multinomial distributions for the original phase-1 data conditional on the phase-2 data. Then we generated synthetic phase-1 data from these estimated distributions conditional on the synthetic error-free data.
    \end{enumerate}
    \item We created synthetic data for the asthma endpoint in a similar manner, although in the original data, only a subset of those in the obesity study were included in the asthma study. We generated synthetic data in a similar manner. Specifically,
    \begin{enumerate}
        \item To simulate which records were included in the asthma study, we included records in a manner similar to that seen in the original data such that the number included exactly equaled that of the original sampling frame, 7053. This was achieved by including all records with synthetic phase-1 time-to-obesity (censored or uncensored) between 4 and 6 years, by sampling a representative proportion of those with synthetic phase-1 time-to-obesity exactly equal to 6 years, and by sampling representative proportions of those with synthetic phase-1 time-to-obesity less than 4 years with and without synthetic phase-1 obesity.
        \item We fit a logistic regression model to the original phase-2 data predicting child asthma status based on estimated weight change, BMI, maternal age, race, ethnicity, diabetes status, maternal asthma, cesarean section, male sex, tobacco use during pregnancy, insurance status, and gestational age. This model was weighted using the appropriate sampling probabilities and calibrated using generalized raking techniques described in this manuscript.
        \item We generated error-free synthetic child asthma status from the fitted logistic regression model in (b) by applying it to the error-free synthetic data and sampling from a Bernoulli distribution based on the fitted probabilities. We increased the frequency of asthma in the synthetic data by adding a constant value of 1 to the linear predictor prior to sampling.
        \item We fit a logistic regression model to predict the original phase-1 child asthma status conditional on phase-2 child asthma status, maternal asthma status, ethnicity, race, sex, and estimated gestational age (EGA). 
        \item Synthetic phase-1 child asthma status was then generated from a Bernoulli distribution with probability of success calculated from the fitted model in (d) based on synthetic error-free data.
    \end{enumerate}
    \item Create phase-2 sampling strata for the synthetic data, similar to those in the original study based on the synthetic phase-1 data. In the original study, we had 33 strata for the obesity study and 10 separate strata for the asthma study. For simplicity, we only created 33 strata for the obesity study in our synthetic data. These synthetic strata were created using the exact same boundaries with the synthetic phase-1 data as were used to create strata in the original study. Although the strata boundaries were identical, the number of records in each stratum were not the same between the original and synthetic strata.
    \item Randomly sample the phase-2 subset from the synthetic data strata, with numbers comparable to the numbers sampled from each stratum in the original study. 
    \item Delete synthetic $(X,Y)$ for those records that were not sampled into the synthetic phase-2 subset, such that $(X,Y)$ are only available on the $n=996$ records in the synthetic phase-2 subset.
\end{enumerate}

\newpage

\startsupplement
\subsection*{S2 Supplemental tables}
\begin{table}[ht]
\caption{Regression coefficients (standard errors) for weight gain during pregnancy comparing multiple imputation with \texttt{mice} and Rubin's rules versus the bootstrap with \texttt{bootImpute} with 200 bootstrap replications and 2 imputations within each bootstrap.}
\begin{center}

\begin{tabular}{rrr}
  \hline
Coefficient & MI Rubin's rules & MI Bootstrapping \\ 
  \hline
Intercept & 0.503 (0.021) & 0.498 (0.020)\\ 
  BMI$^*$ & -0.015 (0.003) & -0.014 (0.003)\\ 
  Age$^*$ & -0.032 (0.006) & -0.031 (0.006)\\ 
  Asian & -0.036 (0.016) & -0.031 (0.017)\\ 
  Black & -0.032 (0.007) & -0.035 (0.010)\\ 
  Other & -0.020 (0.010) & -0.022 (0.018)\\ 
  Hispanic & 0.006 (0.013) & 0.004 (0.014)\\ 
  Depression & 0.038 (0.024) & 0.038 (0.020)\\ 
  Public Insurance & -0.040 (0.015) & -0.038 (0.014)\\ 
  Smoking & -0.032 (0.020) & -0.039 (0.017)\\ 
   \hline
\end{tabular}
\end{center}
\footnotesize{* Body Mass Index (BMI) coefficient is per 5 kg$/m^2$; Age coefficient is per 10 years}

\label{tab:weight-gain-boot}
\end{table}
\clearpage
\newpage

\begin{table}[ht]
\caption{Regression coefficients (standard errors) for logistic regression model of childhood asthma comparing multiple imputation with \texttt{mice} and Rubin's rules versus the bootstrap with \texttt{bootImpute} with 200 bootstrap replications and 2 imputations within each bootstrap.}
\begin{center}

\begin{tabular}{rrr}
  \hline
Coefficient & MI Rubin's rules & MI Bootstrapping \\ 
  \hline
Intercept & -1.846 (2.072) & -1.967 (1.602)\\ 
Weight Change & 1.153 (0.459) & 1.110 (0.490)\\
  BMI$^*$ & 0.173 (0.051) & 0.170 (0.064)\\ 
  Age$^*$ & -0.271 (0.163) & -0.254 (0.148)\\ 
  Black & 1.100 (0.165) & 1.129 (0.216)\\ 
  Other & 0.655 (0.270) & 0.689 (0.280)\\ 
  Male Sex & 0.764 (0.195) & 0.761 (0.202)\\
  Public Insurance & 0.358 (0.268) & 0.419 (0.268)\\ 
  Smoking & 0.373 (0.344) & 0.431 (0.319)\\ 
  Maternal Asthma & 0.578 (0.294) & 0.612 (0.358)\\ 
  EGA$^*$ & -0.046 (0.046) & -0.045 (0.043) \\
   \hline
\end{tabular}
\end{center}

\footnotesize{* Body Mass Index (BMI) coefficient is per 5 kg$/m^2$; Age coefficient is per 10 years; EGA: estimated gestational age}
\label{tab:asthma-boot}
\end{table}

\end{document}